\documentclass[12pt]{iopart}
\usepackage{graphicx,epsf}
\usepackage{color}
\usepackage{citesort}

\usepackage{iopams}
\usepackage{bm}

\usepackage{float}

\newcommand{\ve}{\varepsilon}

\newcommand{\red}{\textcolor{black}}

\begin{document}

\title{Coulomb interaction effects on the Majorana states in quantum wires}


\author{Andrei Manolescu$^1$, D C Marinescu$^{2,3}$ and Tudor D Stanescu$^4$}
\address{$^1$ School of Science and Engineering, Reykjavik University, Menntavegur 1,
IS-101 Reykjavik, Iceland}
\address{$^2$ Department of Physics and Astronomy, Clemson University, Clemson, SC 29634}
\address{$^3$ Kavli Institute for Theoretical Physics, University of California, Santa Barbara, California 93106-4030, USA}
\address{$^4$ Department of Physics and Astronomy, West Virginia University, Morgantown, WV 26506 }
\ead{manoles@ru.is}

\begin{abstract}
The stability of the Majorana modes in the presence of a repulsive
interaction is studied in the standard semiconductor wire -
metallic superconductor configuration. The effects of short-range
Coulomb interaction, which is incorporated using a purely repulsive
$\delta$-function to model the strong screening effect due to the
presence of the superconductor,  are determined within a  Hartree-Fock
approximation of the  effective  Bogoliubov-De Gennes Hamiltonian
that describes the low-energy physics of the wire. Through a numerical
diagonalization procedure we obtain interaction corrections to the single
particle eigenstates and calculate the extended topological phase diagram
in terms of the chemical potential and the Zeeman energy.  We find that,
for a fixed Zeeman energy, the interaction shifts the phase boundaries
to a higher chemical potential, whereas for a fixed chemical potential
this shift can occur either to lower or to higher Zeeman energies. This
effects can be interpreted as a renormalization of the g-factor
due to the interaction.  The minimum Zeeman energy needed to realize
Majorana fermions decreases with increasing the strength of the Coulomb
repulsion. Furthermore, we find that in wires with multi-band occupancy
this effect can be enhanced by increasing the chemical potential,
i. e. by occupying higher energy bands.
\end{abstract}

\maketitle

\section{Introduction}

In the quest for a topological quantum
computer \cite{Kitaev2003,Nayak2008}, the intense investigation of
topological phases of matter \cite{Wilczek2009,Hasan2010,Qi2011} has
propelled the elusive Majorana fermion \cite{Majorana1937} to the
center of a variety of physical systems that enables the exploitation
of its non-Abelian properties \cite{Ivanov2001,Read2009}. One
of the most promising schemes for engineering the topological
superconducting phase that hosts zero-energy Majorana  bound
states \cite{Read2000,Kitaev2001,DSarma2006,Fu2008} involves a
semiconductor (SM) with strong spin-orbit coupling in the presence of an
applied Zeeman field and proximity-coupled to an s-wave superconductor
(SC) \cite{Sau2010,Alicea2010,Lutchyn2010,Oreg2010}.

The recent experimental realization of the one-dimensional (1D) version of
this proposal \cite{Lutchyn2010,Oreg2010}
in semiconductor-superconductor hybrid structures -- the so-called Majorana wire --
has generated reports of
signatures consistent with the presence of zero-energy Majorana  bound
states, such as the fractional Josephson effect \cite{Rokhinson2012}
and the emergence of a zero-bias peak in the differential
conductance \cite{Mourik2012,Das2012,Deng2012,Churchill2013,Finck2013}.
However, these encouraging experimental advances have also underscored
a series of discrepancies between predicted and observed features, as
well as potential problems \cite{Stanescu2013a}, most notably the soft
gap that characterizes the proximity-induced superconductivity in SM
nanowire-SC hybrid structures \cite{Mourik2012,Takei2013,Stanescu2013b},
the absence of any signature associated with the closing of the
quasiparticle gap \cite{Mourik2012,Stanescu2012} at the topological
quantum phase transition (TQPT) \cite{Read2000}. Moreover, the possibility
that observable features similar to those generated by Majorana
bound states can appear in the topologically-trivial phase due to, e.~g.,
strong disorder \cite{Liu2012,Bagrets2012,Pikulin2012,PhysRevB.87.024515}, soft confining
potentials \cite{Kells2012}, or Kondo physics \cite{Lee2012} has not been clearly eliminated.
\red {Zero-bias anomalies have been recently related to parity crossings
of Andreev levels and regarded as precursors of Majorana modes in short 
nanowires \cite{Lee2013}.  }

In general,
while the Majorana bound states are expected to emerge in a topological SC
phase and, consequently, to enjoy a certain degree of protection against
small perturbations, the stability of the Majorana mode as
well as some of the observable features that it generates
\cite{Mourik2012,Das2012,Deng2012,Churchill2013,Finck2013} depend
critically on certain details of the system, such as disorder, multiband
occupancy, finite size effects \cite{Lim2012,Serra2013}, barrier
potentials, and strength of the effective coupling between the SM wire and
the SC and between the wire and the metallic lead \cite{Stanescu2013a}. In this context, it becomes
critical to incorporate the effects of the Coulomb interaction
using approaches that are both reliable and simple enough to accommodate
these influences.

Electron-electron interactions are expected to affect the stability of
the Majorana modes
\cite{Gangadharaiah2011,Sela2011,Lutchyn2011b,Stoudenmire2011,Lobos2012,Hassler2012,PhysRevB.88.161103}
by renormalizing the induced SC pairing
potential and the characteristic length scale of the zero-energy
bound states \cite{Gangadharaiah2011}.  Under strong interaction, a complete suppression of the induced SC pair
potential can occur\cite{Gangadharaiah2011,Sela2011}, along with an increase in the
localization length of the Majorana modes \cite{Stoudenmire2011}. The
interplay between disorder and interaction in Majorana wires was also predicted to favor a quantum phase transition from a
topological SC phase to a topologically-trivial localized phase \cite{Lobos2012}.
At the same time, the presence
of interactions was found to broaden the chemical potential range that supports
the topological SC phase \cite{Stoudenmire2011}, thus  enhancing the
immunity of the Majorana modes against local fluctuations of the chemical
potential.
\red{A broadening of the chemical potential window for Majorana modes has also been
predicted for a different physical setup, based on an array of superconducting islands 
with mutual interaction \cite{Hassler2012}.}

In nanowires, the effect of
interactions is enhanced by the the effective low dimensionality of the
system. Consequently, it was found that in multiband nanowires the phase boundaries
in the Zeeman field -- chemical potential plane are renormalized \cite{Lutchyn2011b}. Further, the transport properties of a nanowire having a structure similar
to that used in the recent experiments \cite{Mourik2012}, i.~e., with part
of the wire's length proximity-coupled to the SC and a normal segment
serving as a lead, have been studied \cite{Fidkowski2012} by treating the
system as a superconductor-Luttinger liquid junction. Finally, repulsive
interactions were recently predicted \cite{Haim2013} to drive a mechanism
that results in the reversal of the sign of the effective pair potential
in the wire and the emergence of time-reversal invariant topological
superconductivity \cite{Wong2012,Nakosai2012,Deng2012b,Seradjeh2012,Zhang2013,Nakosai2013,Keselman2013,Liu2013,Gaidamauskas2013,Klinovaja2013}.

Existing studies have included interaction
effects using a variety of approaches ranging from the Hartree-Fock
approximation, to bosonization and the density-matrix renormalization
group technique. While some of these methods can accurately account
for the effects of interactions, they have serious limitations when
key characteristics of the system, such as the finite wire thickness
or various details of the SC proximity effect \cite{Stanescu2013a},
are incorporated into the theoretical description.

In this paper we study a standard Majorana wire configuration -
a semiconductor wire proximity coupled to a superconductor in the
presence of a repulsive $\delta$ function interaction that realistically
describes a strongly screened regime. In the standard many-body manner,
we incorporate this interaction in a Hartree-Fock-type approximation
of the Bogoliubov - de Gennes (BdG) Hamiltonian that can be reliably
used to account for the effects of Coulomb interaction as well as
other experimentally relevant features, such as the finite length of
the wire.  Within this formalism we calculate the Majorana states in a
flat, two-dimensional quantum wire of both finite and infinite length.
We emphasize that this formalism can be naturally expanded to incorporate
finite range interactions and experimentally-relevant conditions, such
as gate-induced and disorder potentials. Our calculations show that in
the finite wire the modes increase their localization length, while for
an infinite wire the phase diagram describing transitions to topological
superconductor phase in terms of chemical potential and Zeeman energy
depends on the strength of the Coulomb interaction. On account of the
interaction, for a fixed Zeeman energy the phase boundaries are pushed
to higher chemical potentials, whereas for a fixed chemical potential
the boundaries can shift either to lower or to higher Zeeman energies.
However, when the Zeeman energy is large enough such
that the spins are fully polarized along the magnetic field, the phase
boundaries do not depend on the interaction strength.

The outline presented above is reflected by the structure of the paper, as we describe the physical model in Section 2, the effective Bogoliubov - de Gennes Hamiltonian which includes the Coulomb
interaction is discussed in Section 3, while the results for the finite and infinite wire are shown in Section 4.
A cumulative summary of conclusions is presented in Section 5.

\section{The Model}

\begin{figure} [H]
\begin{center}
\includegraphics [width=10 cm] {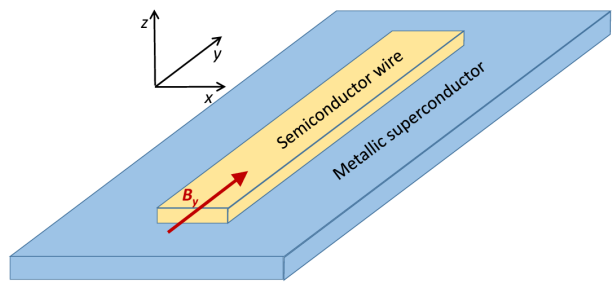}
\end{center}
\caption{The physical system: A semiconductor quantum wire is built on the
surface of a metallic s-wave superconductor. The superconductivity is induced in the
wire by proximity effect.
The wire is quasi two dimensional, with thickness 
$L_z$ much smaller than the width $L_x$, which is much smaller than the length
$L_y$.  A external longitudinal magnetic field $B_y$ is used to create a Zeeman
splitting. }
\label{Sample}
\end{figure}

The following considerations are based on the standard model of a
\red{thin}
Majorana-wire of width $L_x$ and length $L_y$ placed in a magnetic field
$B_y$ oriented along the $y$ direction, i.~e. parallel to the
wire, that induces a Zeeman splitting, \red{as shown in Fig.\ \ref{Sample}.
The wire is strongly confined in the $z$ direction, i.\ e. $L_z \ll L_x$, such
that only the lowest mode in the $z$ direction is relevant for the electronic states.}
The semiconductor (SM) wire is endowed with a Rashba spin-orbit
interaction (SOI) linear in the electron momentum, and of strength
$\alpha$. 

The single particle Hamiltonian of an electron of momentum
$\bf p$ and spin $\sigma$ is consequently written as
\begin{equation}
H = \frac{p^2}{2m}+\sigma_yB + \alpha(\sigma_xp_y-\sigma_yp_x) \ .
\label{ham}
\end{equation}
The boundary conditions are set by hard-walls at
$x=\pm L_x/2$ and $y=\pm L_y/2$. We use the basis generated by the single particle
eigenstates of $H_0=p^2/2m$, labeled by $q$, and described as
$|q\rangle=|n_x n_y s\rangle$, or
$\varphi_q({\bf r}) = \varphi_{n_x}(x)\varphi_{n_y}(y)|s\rangle$, where
$n_x,n_y$ are positive integers, $s$ is the spin projection along $z$, and
\begin{equation}
\hspace{-15mm}
\varphi_{n_x}(x) = \sqrt{\frac{2}{L_x}} \sin \left[ {\frac{n_x(x+L_x/2)\pi}{L_x}}\right ] \ ,
\
\varphi_{n_y}(y) = \sqrt{\frac{2}{L_y}} \sin \left[ {\frac{n_y(y+L_y/2)\pi}{L_y}}\right ] \ .
\label{basis}
\end{equation}
In order to simplify notations the same symbol $\varphi$ is used
for the one-dimensional wave functions corresponding to $x$
and $y$ directions, and for the basis vector as well.

%


When proximity-coupled to a superconductor the wire is described by a many-body
Hamiltonian written in terms of the single-particle creation and destruction
operators $c_q^{\dagger}$ and $c_q$ associated with the basis states
(\ref{basis}) \cite{Lutchyn2011a},
\begin{equation}
\hat K_S = \sum_{q,q'} \left[ \left(H_{qq'}-\mu\delta_{qq'}\right) c_q^{\dagger}c_{q'}^{}+
\frac{1}{2}\left( \Delta_{qq'}c_q^{\dagger}c_{q'}^{\dagger}+H.\ c.\right) \right] \ .
\label{hams}
\end{equation}
In Eq. (\ref{hams}) the proximity-induced superconductivity is described by the pairing potential $\Delta$ with matrix elements
$\Delta_{qq'}=\delta_{n^{}_x n'_x}\delta_{n^{}_y n'_y}\Delta_{s^{}s'}$,
where $|q\rangle=|n_x n_y s\rangle$ and $|q'\rangle=|n'_x n'_y s'\rangle$.
Within the spin space $\Delta_{+1+1}=\Delta_{-1-1}=0$ and
$\Delta_{+1-1}=-\Delta_{-1+1}=\Delta$, which is the pairing energy in the induced
superconductor. $H_{qq'}$ represent the matrix elements of Hamiltonian (\ref{ham}) and 
$\mu$ denotes the chemical potential.

In addition to the attractive pairing induced through the proximity effect,
the electrons in the wire also experience a purely repulsive Coulomb
interaction (\ref{hams}),
\begin{equation}
\hat V = \frac{1}{2} \sum_{q,q',p',p} V_{qq',p'p}
c_q^{\dagger}c_{q'}^{\dagger}c_{p'}^{} c_p^{} \ ,
\label{hamc}
\end{equation}
containing the matrix elements of the Coulomb potential
\begin{equation}
\hspace{-15mm}
V_{qq',p'p}=\langle q q'| u({\bf r - r'}) | p' p\rangle
=\int d{\bf r} d{\bf r'} \varphi^{\dagger}_q({\bf r}) \varphi^{\dagger}_{q'}({\bf r'})
u({\bf r - r'}) \varphi_{p'}({\bf r'})\varphi_p({\bf r}) \nonumber \ .
\end{equation}
On account of the proximity to the superconductor the
Coulomb electron-electron repulsion is expected to be strongly screened, suitably described by
a delta function,
\begin{equation}
u({\bf r - r'})= u_0 \delta(\bf{ r - r'}) \ ,
\label{scp0}
\end{equation}
with $u_0$ a sample specific parameter. As described in the Appendix, $u_0$ is
estimated to be of the order of some eVnm$^2$.
In the reference basis (\ref{basis}),
the matrix elements of the Coulomb potential can be calculated analytically. With
$|q\rangle=|n_x n_y s\rangle$, $|q'\rangle=|n'_x n'_y s'\rangle$,
$|p\rangle=|m_x m_y t\rangle$, $|p'\rangle=|m'_x m'_y t'\rangle$,
we obtain
\begin{equation}
V_{qq',p'p}  = \frac{u_0}{4L_xL_y}\delta_{st}\delta_{s't'}  \
K(n_x,n'_x,m'_x,m_x) \ K(n_y,n'_y,m'_y,m_y) \ ,
\end{equation}
where
\begin{equation}
K(n_1,n_2,m_2,m_1)= \!\!\!\!\!
\sum_{i,j,k=\pm 1} ijk \ \delta_{n_1+im_1,jn_2+km_2}  \ .
\end{equation}

\section{The effective Bogoliubov - de Gennes Hamiltonian}

The spectrum of quantum states of the superconducting wire is found by rewriting
the original Hamiltonian via canonical transformations of the
field operators \cite{Fetter,March}.
Although the derivation of the Bogoliubov - de Gennes Hamiltonian is textbook
material \cite{deGennes}, we prefer to summarize it here for clarity.
The Bogoliubov operators for the creation and destruction of an
excited state with energy $\ve$, $\gamma_{\ve}^{\dagger}$
and $\gamma_{\ve}$, are introduced.  They satisfy Fermionic anticommutation rules i. e.
\begin{equation}
\{ \gamma_{\ve},\gamma_{\ve'}^{\dagger} \} =\delta_{\ve\ve'} \ , \
\{ \gamma_{\ve},\gamma_{\ve'} \} =
\{ \gamma_{\ve}^{\dagger},\gamma_{\varepsilon'}^{\dagger} \} =0 \ ,
\end{equation}
while satisfying the particle-hole symmetry,
$\gamma_{\ve}^{\dagger} = \gamma_{-\ve}$.
Operators $c_q$ are expanded as
\begin{equation}
c_q=\sum_{\ve > 0} \left( u_{\varepsilon q} \gamma_{\varepsilon}
+ v_{\ve q}^*\gamma_{\varepsilon}^{\dagger} \right) \ , \
c_q^{\dagger}=\sum_{\ve > 0} \left( u_{\varepsilon q}^* \gamma_{\varepsilon}^{\dagger}
+v_{\ve q}\gamma_{\varepsilon} \right) \ ,
\label{cantr}
\end{equation}
where $u_{\ve q}\equiv \{ c_q,\gamma_{\varepsilon}^{\dagger} \}$ and
$v_{\ve q}\equiv \{ c_q^{\dagger},\gamma_{\varepsilon}^{\dagger} \}$
are numerical (complex) coefficients to be determined.
Eqs.~(\ref{cantr}) are inserted in the Hamiltonian (\ref{hams}) where
the $\gamma$'s are subsequently rearranged  in the normal order (creation part to the left, destruction
part to the right, and a negative sign for an odd number of permutations), using
Wick's theorem \cite{Fetter}.
In the simplest case of only two operators one obtains, for example,
\begin{equation*}
c_q^{\dagger}c_{q'}^{}=N(c_q^{\dagger}c_{q'}^{})+c_q^{\dagger\cdot}c_{q'}^{\cdot} \ ,
\end{equation*}
and the contraction (second term) can easily be calculated by taking the expectation
value on the vacuum of the $\gamma$'s, $\gamma_{\ve}|0\rangle=0$. The
expected value of the normal product is zero, and hence
\begin{equation*}
\hspace{-15mm}
c_q^{\dagger\cdot}c_{q'}^{\cdot} = \sum_{\ve>0} v_{\varepsilon q} v_{\varepsilon q'}^{*} ,\
c_q^{\cdot}c_{q'}^{\dagger\cdot}=\sum_{\ve>0} u_{\varepsilon q} u_{\varepsilon q'}^{*} ,\
c_q^{\dagger\cdot}c_{q'}^{\dagger\cdot}=\sum_{\ve>0} v_{\varepsilon q} u_{\varepsilon q'}^{*} ,\
c_q^{\cdot}c_{q'}^{\cdot}=\sum_{\ve>0} u_{\varepsilon q} v_{\varepsilon q'}^{*}  .
\end{equation*}
After calculating the normal (N) products the Hamiltonian (\ref{hams}) becomes
\begin{equation}
{\hat K}_S= E_0+\sum_{\ve, \varepsilon' >0}\left[
h_{\ve\varepsilon'}^{(11)} \gamma_{\varepsilon}^{\dagger} \gamma_{\varepsilon'}
+\left(h_{\ve\varepsilon'}^{(20)}\gamma_{\varepsilon}^{\dagger}
\gamma_{\ve'}^{\dagger} + H.\ c.\right)
\right] \ ,
\end{equation}
where the following notations have been used:
\begin{equation}
E_0=\sum_{\ve>0,q,q'}\left[H_{qq'}v_{\ve q}v_{\ve q'}^*+\frac{1}{2}\Delta_{qq'}
(v_{\ve q}u_{\ve q'}^*-u_{\ve q}v_{\ve q'}^*)\right] \ ,
\nonumber
\end{equation}

\begin{equation}
h_{\ve\ve'}^{(11)}=\sum_{q,q'}
\Big[ H_{qq'}(u^*_{\ve q}u_{\ve' q'}-v_{\ve' q}v_{\ve q'}^*)
+ \Delta_{qq'}(u_{\ve q}^*v_{\ve' q'}-v_{\ve q}^*u_{\ve' q'}) \Big] \ ,
\end{equation}
\begin{equation}
h_{\ve\ve'}^{(20)}=\sum_{q,q'}
\Big[ H_{qq'}(u_{\ve q}^*v_{\ve' q'}^*+\frac{1}{2}
\Delta_{qq'}(u_{\ve q}^*u_{\ve' q'}^*-v_{\ve q}^*v_{\ve' q'}^*) \Big] \ .
\end{equation}
$E_0$ is the ground-state energy of the superconductor condensate in the quantum wire.
$h_{\ve\ve'}^{(11)}$
contains the spectrum of excitations and it can be rewritten as
\begin{equation}
h_{\ve\ve'}^{(11)}=\sum_{qq'}
\left( \begin{array}{cc}
u_{\ve q}^*\ , \!\! & v_{\ve q}^*
\end{array} \right)
H_{BdG}
\left( \begin{array}{c}
u_{\ve' q'}\\
v_{\ve' q'}
\end{array} \right) \ .
\end{equation}
$u_{\ve q}$ and $v_{\ve q}$ can be seen as particle/hole (or isospin) components
of a combined wave vector $\Psi_{\ve q}=(u_{\ve q}, v_{\ve q})$, which define the
so-called Nambu space. The central matrix is known as
Bogoliubov - de Gennes (BdG) Hamiltonian
\begin{equation}
H_{\rm BdG}=
\left( \begin{array}{cc}
H_{qq'} & \Delta_{qq'} \\ 
-\Delta_{qq'} & -H_{q'q}
\end{array} \right) \ .
\end{equation}
If $|\Psi_{\ve q}\rangle$ is an eigenvector of $H_{\rm BdG}$ with eigenvalue $\ve$
then one can show \cite{March,Fetter} that $h_{\ve\ve'}^{(20)}=0$.  This means that the
excitation spectrum of the quantum wire in the superconductive state is the spectrum
of $H_{BdG}$.

Similarly, the
two-particle interaction in (\ref{hams}) is subjected to the canonical transformation, leading to four $\gamma$ operator products \cite{March}.  In this case of the Coulomb interaction Wick's theorem gives,
\begin{equation*}
c_q^{\dagger}c_{q'}^{\dagger}c_{p'}^{} c_p^{} = N_4+N_2+N_0 \ ,
\end{equation*}
were $N_{4}$ is the normal product with no contraction, $N_2$ is the sum
of all combinations with only one pair of $\gamma$'s contracted, and $N_0$ is the sum with
two pairs contracted.  $N_0$ is only a c-number and has a contribution
to the energy of the condensate $E_0$.  $N_4$ has combinations of four $\gamma$
operators which can be interpreted as Coulomb correlations of higher order than $N_2$
and will be neglected.  We are thus left with $N_2$,
\begin{equation}
N_2=N_2^{(12)}+N_2^{(13)}+N_2^{(14)}+N_2^{(23)}+N_2^{(24)}+N_2^{(34)} \ ,
\nonumber
\end{equation}
where the upper indices of each term indicate the contracted operators:
\begin{eqnarray}
&&N_2^{(12)}=N(c_q^{\dagger\cdot}c_{q'}^{\dagger\cdot}c_{p'}^{} c_p^{})=
 \sum_{\ve>0} v^{ }_{\ve q } u^{*}_{\ve q'} N(c_{p'}^{}c_p^{}) \ ,\nonumber\\
&&N_2^{(13)}=N(c_q^{\dagger\cdot}c_{q'}^{\dagger}c_{p'}^{\cdot} c_p^{})=
-\sum_{\ve>0} v^{ }_{\ve q } v^{*}_{\ve p'} N(c_{q'}^{\dagger}c_p^{}) \ , \nonumber\\
&&{\rm etc}. \nonumber\\
\end{eqnarray}

After the remaining $N$ products are calculated and inserted in (\ref{hamc}),
the coefficient of $\gamma^{\dagger}_{\ve}\gamma^{}_{\ve'}$ are identified and
added to $h_{\ve\ve'}^{(11)}$. The changes of the BdG Hamiltonian due to the Coulomb
contributions are:
\begin{equation*}
H_{qq'} \to H_{qq'}+W_{qq'} \ , \ \
\Delta_{qq'} \to \Delta_{qq'} + \Gamma_{qq'} \ ,
\end{equation*}
where
\begin{eqnarray}
&& W_{qq'}=\sum_{\ve>0,p,p'} v^{ }_{\ve p } v^{*}_{\ve p'} (V_{qp,p'q'}-V_{qp,q'p'}) \label{W} \\
&& \Gamma_{qq'}=\frac{1}{2} \sum_{\ve>0,p,p'} u^{ }_{\ve p } v^{*}_{\ve p'}
(V_{qq',pp'}-V_{q'q,pp'}) \label{Gamma}
\end{eqnarray}
In the derivation of these results the symmetry properties of the Coulomb
matrix elements have been used: $V_{qq',p'p}=V_{q'q,pp'}=V_{pp',q'q}^*$.
We obtain $W_{qq'}=W_{q'q}^*$ and $\Gamma_{qq'}=-\Gamma_{q'q}$,
in agreement with the similar symmetries of the non-Coulomb term.
On account of the electron-hole symmetry we also write,
\begin{equation*}
\sum_{\ve>0} v^{ }_{\ve p } v^{*}_{\ve p'}=
\sum_{\ve<0} u^{*}_{\ve p } u^{ }_{\ve p'}\nonumber \ ,
\end{equation*}
the restriction to negative $\varepsilon$ corresponding to the occupied
particle states of the original quantum wire in the normal state, i.\
e. below the chemical potential.  The new matrices $W$ and $\Gamma$
depend themselves on the eigenstates of the BdG Hamiltonian.  Therefore,
to solve the new eigenvalue problem, i.\ e. with Coulomb contributions,
an iterative numerical scheme is implemented.

\section{Results}

\subsection{The quantum wire of finite length}

We consider a quantum wire of width $L_x=100$ nm and length $L_y=5000$ nm, 
\red {and neglect the thickness $L_z$, Fig.\ \ref{Sample}}.  
We note that this geometry represents a good approximation for thin nanoribbons, provided only one transverse mode (i.~e. one confinement-induced band associated with the $z$-direction) is occupied. The material parameters are those of InSb:
$m_{\rm eff}=0.016$, $g_{\rm eff}=-50$, $\alpha=20$ meV~nm.  The energy
spectrum of the spin-less, normal wire, without Coulomb interaction, is 
\begin{equation}
E_{n_x n_y}=(\hbar^2\pi^2/2m_{\rm eff})[(n_x/L_x)^2+(n_y/L_y)^2].
\label{En0}
\end{equation}
In the presence of a magnetic field of strength $B$ the Zeeman energy (i. e the 
Zeeman splitting) in meV is $E_Z=|g_{\rm eff}|\mu_B B \approx 2.9\times B{\rm [T]}$. 
(In other papers the Zeeman energy may be defined as $E_Z=\frac{1}{2}|g_{\rm eff}|\mu_B B$.)
The superconductor
parameter is fixed to $\Delta=0.25$ meV.  The BdG Hamiltonian is
diagonalized numerically in the space covered by the basis $|\tilde
q\rangle = |\eta q\rangle =|\eta n_x n_y s \rangle$ with $\eta=\pm 1$
the Nambu quantum number.  In the following examples we use $n_x\le 4$
and $n_y\le 100$, such that the size of the basis is $2\times 100 \times
4 \times 2 = 1600$.

\begin{figure}
\begin{center}
\includegraphics[width=9cm] {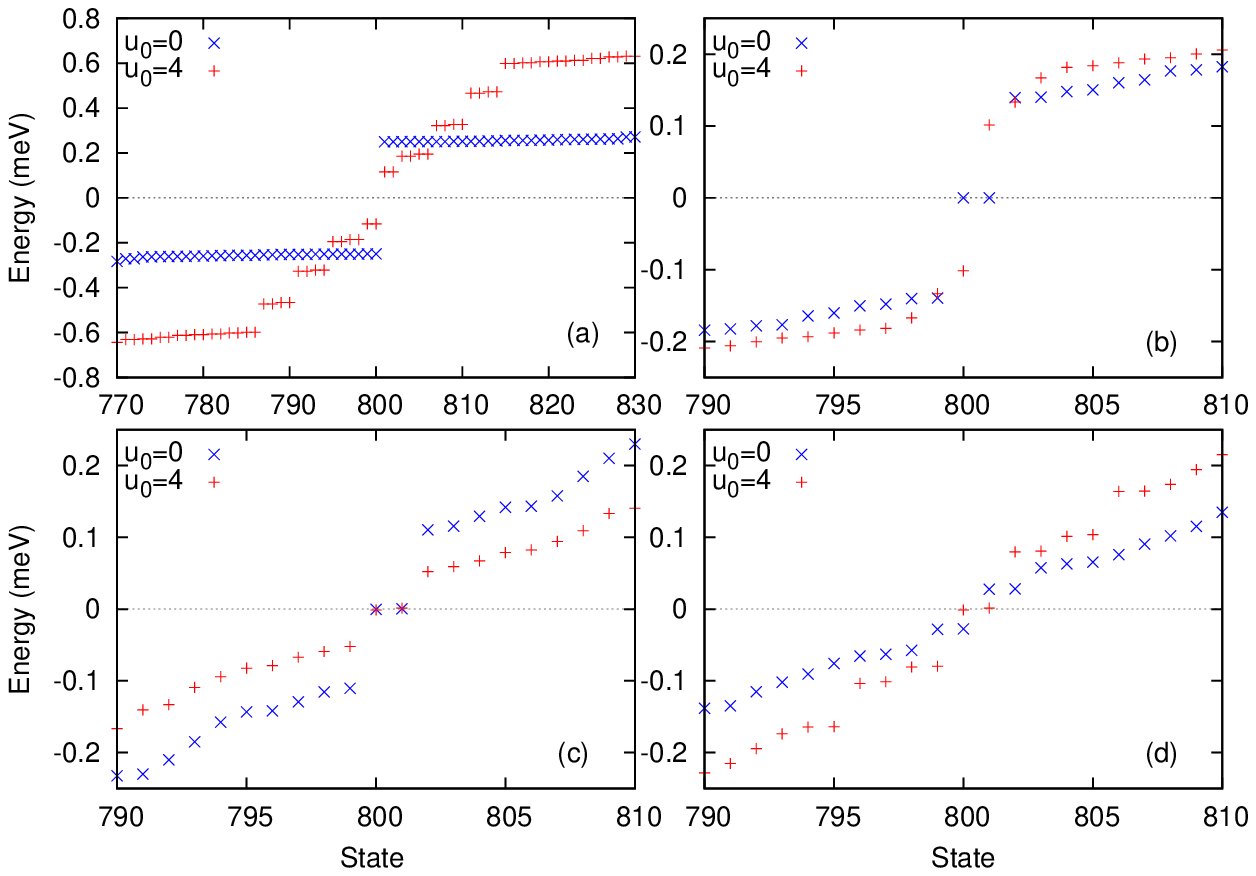}
\end{center}
\caption{The effect of Coulomb interaction of strength $u_0=4\ {\rm eV \ nm^2}$
on the BdG energies for a wire of width $L_x=100$ nm and  finite
length $L_y=5000$ nm for different magnetic fields $B_y$ and chemical potentials $\mu$.
The states are ranked by energy on the horizontal axes.
The zero energy is marked with a dotted line for comparison.
The BdG energies without Coulomb interaction correspond to $u_0=0$.
(a) $\mu=9.4$ meV, $B_y=0$.
(b) $\mu=9.4$ meV, $B_y=0.3$ T ($E_Z=0.87$ meV).
(c) $\mu=9.4$ meV, $B_y=0.68$ T ($E_Z=1.97$ meV).
(d) $\mu=10.6$ meV, $B_y=0.68$ T.  }
\label{spe_finite}
%
\begin{center}
\includegraphics[width=9cm]{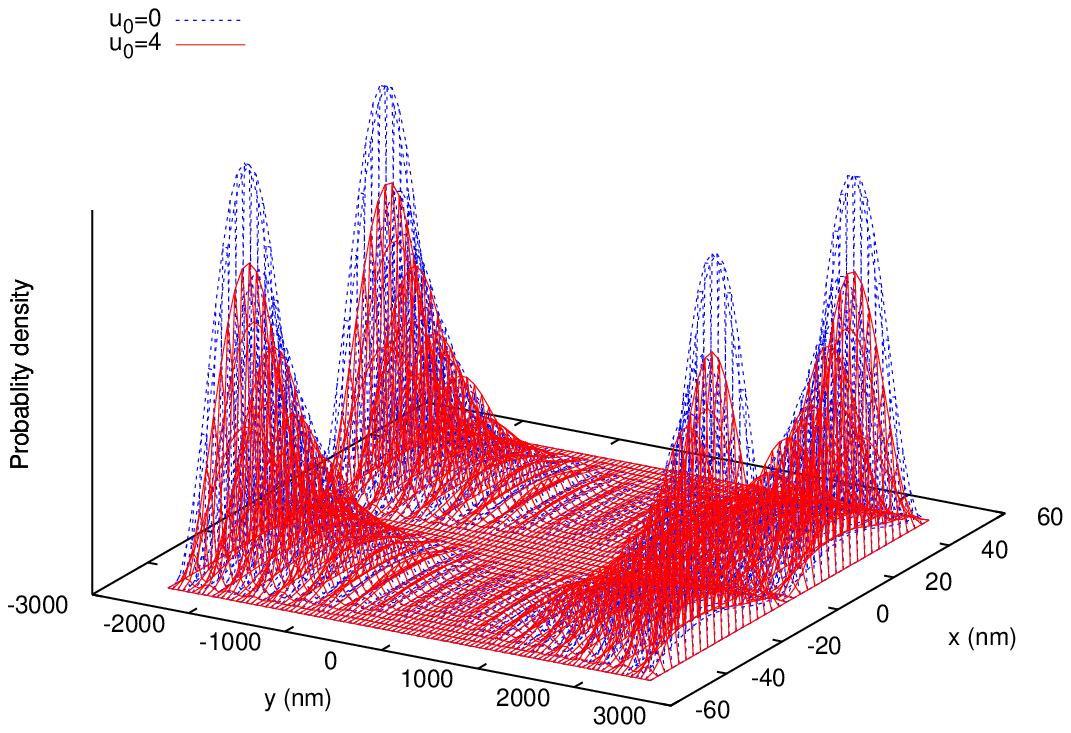}
\end{center}
\caption{Probability densities of corresponding to the particle component of the
Majorana states number 800 of Fig.~\ref{spe_finite}(c), without and with interaction.}
\label{den}
\end{figure}

In Fig.~\ref{spe_finite} we show BdG spectra without and with Coulomb
repulsion.  Only the central part of the spectra is shown in the
figures, corresponding to the low-energy excitations.  In the absence
of a magnetic field and of the Coulomb interaction the induced superconductor
gap of $2\Delta=0.5$ meV is clearly visible in the spectrum shown in Fig.~\ref{spe_finite}(a).
As expected, the gap decreases in the presence of
the Coulomb repulsion, and with $u_0=4 \ {\rm eV \ nm^2}$ the new gap is
0.23 meV. We note that this reduction of the induced SC gap represents a more meaningful measure of the Coulomb effects
than the magnitude of the coupling constant $u_0$ itself.

As the strength of the magnetic field along the wire $B_y$ increases, the superconductor gap disappears at a
certain field value corresponding to a TQPT, 
followed by a reemergence as the system enters
a topological superconductor phase with two Majorana zero-energy states
localized at the two ends of the wire. In the absence of the Coulomb  interaction,
if the Fermi energy is at the bottom of an energy band of the normal
wire, for example $\mu=E_{20}=9.4$ meV, the gap closes when the Zeeman
energy is equal to the induced superconductor gap, 
$E_Z=2\Delta$, i.\ e.\
for $B_y=0.17$~T \cite{Stanescu2011}.  In Fig.~\ref{spe_finite}(b) we
consider a slightly stronger magnetic field, $B_y=0.3$ T, for which the
gap is open again, and two zero-energy Majorana bound states are created in the middle of
the gap.  Within our basis these states are ranked as $800$ and $801$ on
the energy scale. In the interacting case, the system is still in the topologically-trivial phase, but
the quasiparticle gap will close at a higher field $B_y\approx 0.5$ T (not shown).

At an even higher field, for example $B_y=0.68$ T, Majorana
states are obtained both with and without interaction, Fig.~\ref{spe_finite}(c).
The probability distributions of these states are shown for comparison
in Fig.~\ref{den}. As expected, the localization effect at the ends
of the wire is reduced by the interaction, but it is still present and
consistent with states created inside the energy gap. Note that the double peak structure of the Majorana modes is due to the fact that these states are associated with the top occupied band, which, for the parameters used in the calculation, corresponds to $n_x=2$.
Next, in Fig.~\ref{spe_finite}(d) we keep the same magnetic field,
but increase the Fermi energy into the second band, to $\mu=10.6$ meV, i.~e. by $\delta\mu=1.2$ meV.
The results are now opposite to Fig.~\ref{spe_finite}(b), with
the Majorana present in the interacting case, but absent for $u_0=0$, as $E_Z < 2\sqrt{\delta\mu^2 +\Delta^2}$, i.~e. the wire is in the topologically trivial phase.

\subsection{The quantum wire of infinite length}

For the infinite wire the basis functions in the $y$ direction
(\ref{basis}) become plane waves with wave vectors $k$.  For any fixed
$k$, which is here a good quantum number, the eigenstates can be labeled
by $|q\rangle=|n_x s\rangle$.  The BdG Hamiltonian is now diagonal in $k$,
and therefore the matrix elements of the Coulomb terms $W$ and $\Gamma$,
Eqs.\ (\ref{W}-\ref{Gamma}), become $W_{qq'}(k)$ and $\Gamma_{qq'}(k)$.

\begin{figure}
\begin{center}
\includegraphics[width=9cm]{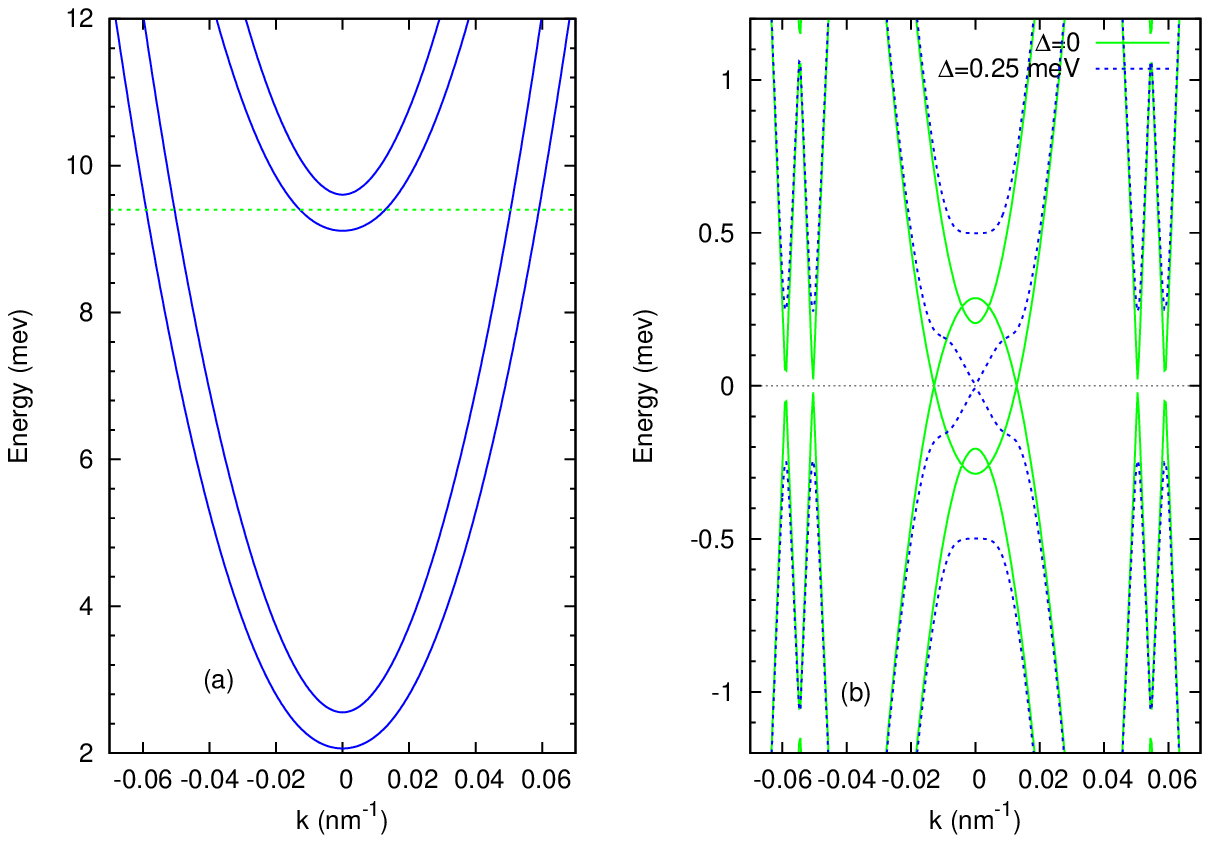}
\caption{Energy spectra for the infinite quantum wire in longitudinal magnetic field
$B_y=0.17$ T, without Coulomb interaction ($u_0=0$).
(a) The normal wire, with the horizontal line showing the Fermi energy $\mu=9.4$ meV.
(b) BdG energies for $\Delta=0$ and $\Delta=0.25$ meV. In the latter case the gap is
closing, indicating the transition to the topological phase with one Majorana state at
each end of the wire.}
\label{spe_inf1}
\end{center}
\begin{center}
\includegraphics[width=10cm]{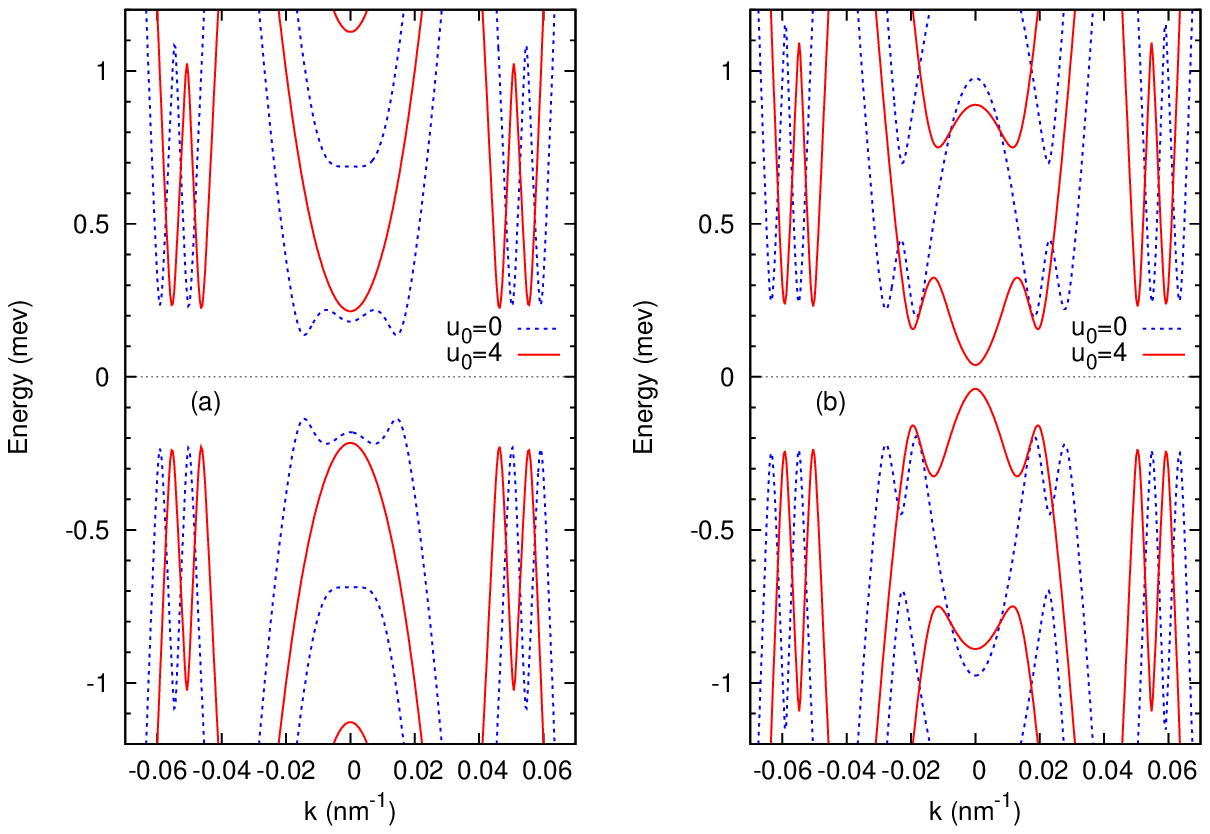}
\caption{BdG spectra for the infinite quantum wire without and with Coulomb
interaction of strength $u_0=4\ {\rm eV \ nm^2}$.
(a) $\mu=9.4$ meV, $B_y=0.3$ T ($E_Z=0.87$ meV), which are the same
parameters as in Fig.~\ref{spe_finite}(b).
(b) $\mu=10.6$ meV, $B_y=0.2$ T ($E_Z=0.58$ meV).  }
\label{spe_inf2}
\end{center}
\end{figure}

In Fig.~\ref{spe_inf1} we show energy spectra for a quantum wire
of infinite length and of width $L_x=100$ nm, without Coulomb
interaction, both in the normal state and at the transition to
the topological superconductor state occurring at minimum magnetic
field, i.\ e.\ for $E_Z=2\Delta$, when the energy gap is closing for
$k=0$ \cite{Stanescu2011}.  For a slightly higher magnetic field, like
$B_y=0.3$ T, the gap reopens and the resulting BdG spectra are shown
in Fig.~\ref{spe_inf2}(a), both with and without Coulomb interaction,
to be compared with the versions obtained for the finite wire and shown
in Fig.~\ref{spe_finite}(b). Consistent with the finite wire calculation, 
for $B_y=0.3$ T the system is in the topological SC phase in the absence of 
interaction, but still in the topologically trivial phase for $u_0=4$ eV~nm.

The gap obtained for the BdG spectrum of the infinite wire with
interaction is 0.43 meV, which is slightly higher than the analogous
result seen in Fig.\ref{spe_finite}(b) for the finite wire of length $5000$
nm, which is $0.20$ meV, i.\ e.\ the energy difference between the states
$800$ and $801$.  Hence the size of the gap with interaction depends on the
length of the wire. Increasing the length of the finite wire to $7000$
nm, while keeping all the other parameters unchanged, we obtain $0.28$ meV.
Increasing the magnetic field for the infinite wire with interaction the
gap of the BdG spectrum decreases.  For example it becomes $0.20$ meV for
$B_y=0.37$ T and nearly zero for $B_y=0.43$ T if $\mu=9.4$ meV (not shown).
The transition can also be
approached by changing the Fermi energy, like in Fig.~\ref{spe_inf2}(b),
where the gap is $0.08$ meV, for $B_y=0.2$ T and $\mu=10.6$ meV. 
The transition of the interacting wire occurs
at a slightly higher field.  This situation is selected to be compared with
Fig.~\ref{spe_finite}(d) which is for a larger field and hence with the
interacting system on the other side of the phase boundary, but with
the noninteracting one still in the trivial phase.

\subsection{$W$ and $\Gamma$ contributions to the BdG spectra}

The Coulomb interaction is included in our model such that the BdG
spectrum is consistent with the electronic spectrum of the system in
the normal state.  The normal state corresponds to  $\Delta=0$, and in
this case the BdG spectrum is the same as for the electrons in the normal
quantum wire, only shifted by the chemical potential, and replicated for
negative energies via particle-hole symmetry, $\varepsilon\rightarrow
-\varepsilon$.  Indeed this is true without interaction (as shown in
Figure \ref{spe_inf1}), since the BdG Hamiltonian becomes block-diagonal
in the Nambu space. In the presence of the interaction, the BdG
Hamiltonian is still diagonal if $\Delta=0$.  The reason is that the
contribution to the off-diagonal terms $\Gamma_{qq'}$ contains products of
$u$ and $v$ wave functions, which for $\Delta=0$ describe pure electrons
or pure holes, and hence one of them is zero.  The numerical calculations
are based on iterations with the initial solution corresponding
to the noninteracting case, i.\ e. with $\Gamma_{qq'}=W_{qq'}=0$,
and if $\Delta=0$ then $\Gamma_{qq'}=0$ for each iteration.  Only the
(Nambu-)diagonal Coulomb term $W_{qq'}$ contributes to the spectrum in
the normal state with interaction, which corresponds to pure electrons
and holes essentially in the Hartree-Fock approximation.  $W_{qq'}$
describes the particle-particle or hole-hole repulsion.

\begin{figure}
\begin{center}
\includegraphics[width=9cm]{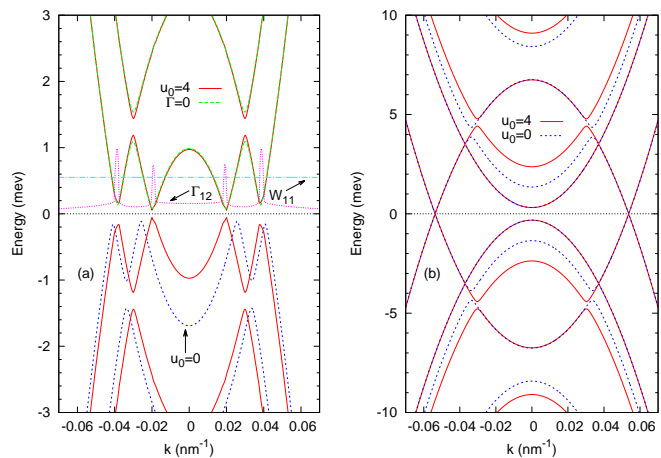}
\caption{BdG spectra for the infinite quantum wire with $\mu=5$ meV.
The solid (red) lines show the results obtained with Coulomb interaction of
strength $u_0=4\ {\rm eV \ nm^2}$ and the (blue) dotted lines those without
interaction, $u_0=0$.
(a) $B_y=0.7$ T ($E_Z=2.03$ meV). The results without interaction are
shown only in the negative (hole) part of the spectrum, for comparison.
In the positive part of the spectrum we compare the results with full
Coulomb interaction with the results obtained when only the $W$ term
is used and $\Gamma$ is neglected ($\Gamma=0$).  The matrix elements $W_{11}$
(independent of $k$) and $\Gamma_{11}(k)$ are also indicated.
(b) $B_y=2.8$ T ($E_Z=8.1$ meV). In this case energy gap at $k=0$ does
not depend on the interaction strength.}
\label{spe_inf3}
\end{center}
\end{figure}

The term $\Gamma_{qq'}$ corresponds to the particle-hole interaction
and it is in general small for the parameters that we consider
in this work.  In the spectra shown for the quantum wire of finite
length the contribution of this term is of the order of 10\%.  For the
infinite quantum wire it is lower and even negligible relatively to the
$W$ term.  Because the matrix elements $\Gamma_{qq'}(k)$ depend on the
particle-hole mixing, $u_a v_a^*$, Eq.\ (\ref{Gamma}), they have peaks
for the wave vectors where energy gaps are small.  An example is shown
in Fig. \ref{spe_inf3}(a).  The net effect of $\Gamma$ is however very
small, as shown in this figure for positive energies, where the spectrum
with the full interaction included is compared with the spectrum where
only the $W$ term is used.  In the same figure, for negative energies, we
compare the spectrum with the full interaction with the spectrum without
interaction($u_0=0$).  So in general, in our study the effects of the
interaction on the BdG spectrum are mostly determined by particle-particle
and hole-hole interactions, and less by particle-hole interactions.

Furthermore, the effects of the Coulomb interaction decreases when the
magnetic field increases.  This is seen in Fig. \ref{spe_inf3}(b), for a
magnetic field of 2.8 T.  The energy spectrum of the normal wire can be
easily distinguished as the quasi-parabolic convex bands with the Fermi
level at zero energy.  In this case all the occupied states have ``spin
up'',  i.\ e.  along the magnetic field  (ignoring the tilting
effect of the SOI).  The Zeeman energy is larger than the orbital energy
and thus the first unoccupied band has the same spin-up orientation and
the topological phase transition occurs when the band bottom is zero.
As long as only spin-up states are occupied the interaction does not
affect the spin-up, but only the spin-down states.  To understand
that we can simplify the normal wire to a single orbital band $E_1(k)$
(i.\ e. $n_x=1$), neglect the SOI ($\alpha=0$), and consider $u_{\ve
q}$ eigenstates of $\sigma_y$.  The interaction term, Eq.\ (\ref{W}),
becomes $W_{ss'}=w_1\delta_{ss'}-w_2(\sigma_y)_{ss'}$, and the interacting
energies are $E_1(k)+w_1\pm (E_Z/2+w_2)$.  If only spin-up states are
occupied then $w_1=w_2$ and the lowest energy is not affected.  The
argument can be extended for two or more bands ($n_x>1$) in a straight
forward manner.  In fact the result is consistent with the Hubbard
model used in Ref. \cite{Stoudenmire2011}, because the interaction has
no effect on the many-body groundstate if only one spin state is occupied.

\subsection{The phase diagram}

\begin{figure}
\begin{center}
\includegraphics[width=9cm]{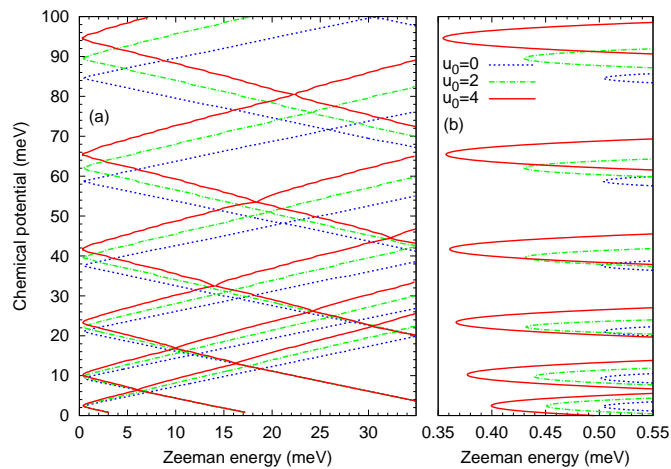}
\caption{(a) The phase diagram in Zeeman energy ($E_Z$) - chemical potential
($\mu$) coordinates.  The phase frontiers for different interaction strength,
$u_0=0,\ 2, \ {\rm and} \ 4 \ {\rm eV \ nm^2}$, are shown with different line types.
The weak oscillations of the phase boundaries are generated by numerical noise.
(b) The corners of the phase frontiers at low Zeeman field are magnified ten times
for visibility, while the relative distance between corners is on the original scale.
Due to the interaction the minimum Zeeman energy needed for the topological transition
decreases when the chemical potential increases.
}
\label{phase_diagram}
\end{center}
\end{figure}

The two models of quantum wire, with finite or infinite length, give us
complementary information on the TQPT and  on the presence of Majorana states.  In a 
finite wire, the Majorana modes localized at opposite ends overlap and, as a result, acquire finite energy. This energy vanishes exponentially in the long wire limit. However, it is rather difficult to describe the TQPT, technically a property of a system with $L_y\rightarrow\infty$,  based on finite wire calculations. Therefore, the phase transitions are described using the infinite wire model and identified by the vanishing of the quasiparticle gap  at $k=0$ \cite{Stanescu2011}. Nonetheless, our numerical calculations show that the boundary between
the trivial and topological superconductor phases, when $\mu$ and
$B_y$ are varied, is almost the same for the finite and infinite wires,
both with and without the Coulomb interaction, as also indicated by the
previous compared examples.  To see the effect of the Coulomb interaction
on the phase diagram, for a larger set of parameters, we use the infinite
quantum wire and define the phase boundaries as the $(E_Z,\mu)$ values
which minimize the gap of the BdG spectrum at $k=0$. The results are
shown in Fig.~\ref{phase_diagram}, for the noninteracting case and for
two different interaction strengths.

The phase boundaries are practically defined by the parameters
corresponding to the chemical potential touching the bottom of a
confinement-induced band in the normal state spectrum of the quantum
wire \cite{Stanescu2011}. For the noninteracting case these boundaries
can be approximated by the linear equations
\begin{equation}
\mu=E_{n_x0}\pm E_Z/2=2.35n_x^2\pm E_Z/2 \ {\rm meV}, 
\label{phb}
\end{equation}
where $n_x=1,2,...$ is the orbital band index, Eq.\ (\ref{En0}).  The effect of the Coulomb
interaction can be understood in a simple manner,   
\red{consistent with our mean-field approximation, i.\ e. neglecting the 
correlation term $N_4$}.  Because of the
repulsive nature, the interaction has a positive contribution to the
energy spectrum of the normal wire, for any given chemical potential and
Zeeman energy.  With all band bottoms moving up in energy, a positive
energy shift is needed for the chemical potential to approach the phase
transition.  Therefore, for any fixed $E_Z$ (i.\ e. a fixed magnetic field)
all phase boundaries move up on the $\mu$ scale because of the interaction,
except when the system is completely spin polarized. 
In the absence of Coulomb interaction, according to Eq.\ (\ref{phb}), this 
occurs when $E_Z>2(\mu-E_{10})$.  In this case the interaction has no effect, 
on the low-energy states of the BdG Hamiltonian as illustrated in Fig.\ \ref{spe_inf3}(b).

Our results are consistent with those of Stoudenmire et al.  obtained
for a strictly one-dimensional model \cite{Stoudenmire2011}.  In our
phase diagram this is the subdomain corresponding to $n_x=1$, i.\ e.
the bottom-left corner of Fig. \ref{phase_diagram}(b), or approximately
$\mu<6, E_Z< 7$\ meV.  As those authors pointed out, within this subdomain
the effect of the interaction is to widen the topological phase at a
fixed $E_Z$ \cite{Stoudenmire2011}.  This is not generally true when more
subbands are involved. As can be seen in Fig. \ref{phase_diagram}(a),
at higher chemical potentials the phase frontiers can shift either to
lower or to higher Zeeman fields in the presence of the interaction.
The multicritical points, i.\ e.\ the intersections of the phase
frontiers, always shift to lower Zeeman fields, more and more in
the higher bands.  The increase of the slope of the phase boundaries
can be understood physically as a result of an interaction-induced
renormalization of the effective $g$-factor. This renormalization,
which is formally due to the off-diagonal spin contributions to $W_{q
q^\prime}$ in Eq. (\ref{W}), depends on the occupancy of the wire, i.~e. on
the chemical potential.

At low Zeeman energies the phase diagram has another interesting
feature. The corners of the phase boundaries split more and more in the
higher bands and the minimum Zeeman energy decreases when the interaction
strength increases, as illustrated in Fig. \ref{phase_diagram}(b).
Therefore, by measuring the the minimum values of the Zeeman field
associated with the low-energy subbands one can experimentally probe
the strength of the interaction effects. Note that, in a noninteracting
wire, these minimum values are the same for all subbands, $E_Z^{(min)}
= 2\Delta$.

\section{Conclusions}

We have considered corrections to the Bogoliubov-De Gennes Hamiltonian
due to short-range repulsive interactions between electrons.  \red {The
short range interaction is justified by the the screening effect of the
superconductor.} Our approach corresponds to a Hartree-Fock approximation
and neglects higher order correlations.  \red {As a reality check we
mention that a similar approach has been recently used to describe
zero-bias anomalies in short quantum wires, in good agreement with
experiments \cite{Lee2013}.  }
We have calculated the energy states numerically, with
an iterative scheme, for finite and infinite models of quantum wires.
We have shown that Majorana states are robust to the Coulomb repulsion.
For the parameters that we used  the interaction effects are dominated
by particle-particle and hole-hole interactions described by the $W$
matrix, Eq.~(\ref{W}). The particle-hole interaction, described by the
$\Gamma$ matrix, Eq.~(\ref{Gamma}), is in general small or negligible.
We have built the phase diagram corresponding to topological phases
containing Majorana modes using the model of the infinite quantum wire
with several bands. The interaction has no effect on the phase frontiers
in fully spin polarized states, i. e.  for a Zeeman energy larger than the
chemical potential, a result which is consistent with a Hubbard model.
Outside this domain the interaction shifts the phase frontiers. The
minimum Zeeman energy needed to generate Majorana fermions decreases
with increasing the strength of the Coulomb repulsion. Moreover, this
effect can be enhanced by increasing the chemical potential to values
corresponding to higher energy confinement-induced bands. We propose
the measurement of the subband dependence of the minimum Zeeman field
as a possible way to experimentally probe the strength of the Coulomb
interaction in Majorana wires.

\

\bibliography{BdGC}
\bibliographystyle{iopart-num}

\ack
This work was supported by the Icelandic Research Fund Grant 10000802, 
National Science Foundation Grant No. PHY11-25915, and
the WV Higher Education Policy Commission Research Challenge Grant HEPC.dsr.12.29.

\section*{Appendix}
\section {Determining the Coulomb coupling constant $u_0$}

A formal justification of  Eq.\ (\ref{scp0}) as well as an estimate for the parameter $u_0$
can be obtained in two different, complementary models. First, we consider a model where
each electron in the semiconductor wire creates an image charge in the superconductor metal. If
$L_z$ is the thickness of the quantum wire ($L_z << L_x << L_y$), then
the distance between the electron and the image charge is of the order
of $L_z$.  Any other electron will see the original negative charge of the
first one in the company of the reflected positive charge.  Therefore a
model of an effective screened Coulomb potential (energy) can be
\begin{equation}
u({\bf r - r'})=
\frac{e^2}{\kappa}\left(\frac{1}{|{\bf r - r'}|}
-\frac{1}{\sqrt{|{\bf r - r'}|^2+L_z^2}}\right) \ ,
\label{scp_im}
\end{equation}
where $\kappa$ is the dielectric constant of the semiconductor host.
For a small $L_z$, in the spirit of the theory of distributions, this
potential is a precursor of a $\delta$-Dirac function, and
Eq.\ (\ref{scp_im}) becomes
\begin{equation}
u({\bf r - r'})=\frac{e^2}{\kappa} (2\pi L_z) \delta({\bf r - r'}) \ ,
\label{scp1}
\end{equation}
the prefactor being calculated by integrating a well behaved function, and
hence $u_0=2\pi e^2 L_z/\kappa$.
In physical terms a small $L_z$ corresponds to a thickness of the quasi two-dimensional electron
gas much smaller than the average inter-electronic distance, i.\ e.\
$L_z \ll \lambda_F$.  For InSb material $\kappa\approx 18$ and $L_z=10 \ {\rm nm}$
with one obtains $u_0\approx 5 \ {\rm eV nm^2}$.

In a complementary approach, we 
 assume that the electrons are localized on atomic lattice sites denoted as
$i \ {\rm or} \ j$.  The repulsive potential is $u_{ij}=V_0\delta_{ij}$,
which in the continuous limit becomes $u(r)=V_0a^3\delta({\bf r})$ in 3D
and $u(r)=V_0a^2\delta({\bf r})$ in 2D.  $a$ is the lattice constant which
for InSb is 0.65 nm.
The energy parameter $V_0$ can be roughly estimated by integrating the Coulomb
potential $1/r$ in 3D on a sphere of radius $a$ and equating the result with
$V_0a^3$, which gives $V_0=2\pi e^2/a$.  Then the 2D prefactor of the $\delta$
potential, Eq.\ (\ref{scp0}), is $u_0=V_0a^2\approx 5.9  \ {\rm eV nm^2}$.

We can refine the lattice method by assuming the electrons described
by atomic orbitals with a constant probability density within a sphere
of radius $a$, which is $3/4\pi a^3$. The electric field inside this uniformly
charged sphere is $e r/a^3$, with $0<r<a$.  The energy of a second electron within
this sphere is $\Phi(r)=e (3a^2-r^2)/2a^3$, where we included the energy of the electron
to sit in the center of the sphere, i.\ e. at $r=0$.  The energy of the pair of electrons
is found by the 3D integration on a sphere of radius $a$ ($S_a$),
\begin{equation*}
V_0=e \frac{3}{4\pi a^3} \int_{S_a} \Phi(r) \ d {\bf r}=\frac{6}{5}\frac{e^2}{a} \approx 2.7 \ {\rm eV} \ .
\end{equation*}
The desired 2D parameter is now $u_0=V_0 a^2\approx 1.1 \ {\rm eV nm^2}$.

\

In conclusion the interaction parameter $u_0$ in Eq.\ (\ref{scp0}) should be of the order of
${\rm eV nm^2}$.

\end{document}